\documentclass[
aps,prd,
showpacs,twocolumn,notitlepage,
amssymb,amsmath,amsfonts,mathrsfs,
nofootinbib,superscriptaddress,
floats,floatfix
]{revtex4-2}

\usepackage{graphicx}
\usepackage{xcolor}
\usepackage[colorlinks=true]{hyperref}
\hypersetup{citecolor=cyan,linkcolor=magenta}
\usepackage{multirow,array}
\DeclareMathAlphabet{\pazocal}{OMS}{zplm}{m}{n}
\usepackage{journals}

\begin{document}
	\title{Dynamical Scalarization during Neutron Star Mergers in scalar-Gauss-Bonnet Theory}
	
	\date{\today}
	
	\author{Hao-Jui Kuan}
	\email{hao-jui.kuan@aei.mpg.de}
	\affiliation{Max Planck Institute for Gravitational Physics (Albert Einstein Institute), 14476 Potsdam, Germany}
	\affiliation{Theoretical Astrophysics, Eberhard Karls University of T\"ubingen, T\"ubingen 72076, Germany}
	
	\author{Alan Tsz-Lok Lam}
	\affiliation{Max Planck Institute for Gravitational Physics (Albert Einstein Institute), 14476 Potsdam, Germany}
	
	\author{Daniela D. Doneva}
	\affiliation{Theoretical Astrophysics, Eberhard Karls University of T\"ubingen, T\"ubingen 72076, Germany}
	\affiliation{INRNE - Bulgarian Academy of Sciences, 1784  Sofia, Bulgaria}
	
	\author{Stoytcho S. Yazadjiev}
	\affiliation{Theoretical Astrophysics, Eberhard Karls University of T\"ubingen, T\"ubingen 72076, Germany}
	\affiliation{Department of Theoretical Physics, Faculty of Physics, Sofia University, Sofia 1164, Bulgaria}
	\affiliation{Institute of Mathematics and Informatics, 	Bulgarian Academy of Sciences, 	Acad. G. Bonchev St. 8, Sofia 1113, Bulgaria}
	
	\author{Masaru Shibata}
	\affiliation{Max Planck Institute for Gravitational Physics (Albert Einstein Institute), 14476 Potsdam, Germany}
	\affiliation{Center of Gravitational Physics and Quantum Information, Yukawa Institute for Theoretical Physics, Kyoto University, Kyoto, 606-8502, Japan} 
	
	\author{Kenta Kiuchi}
	\affiliation{Max Planck Institute for Gravitational Physics (Albert Einstein Institute), 14476 Potsdam, Germany}
	\affiliation{Center of Gravitational Physics and Quantum Information, Yukawa Institute for Theoretical Physics, Kyoto University, Kyoto, 606-8502, Japan} 
	
	\begin{abstract}
		In certain classes of the scalar-Gauss-Bonnet theory strong spacetime curvature in the vicinity of neutron stars and black holes can spontaneously trigger scalarization in the compact object if the coupling strength of the scalar field to the Gauss-Bonnet invariant exceeds a critical value. Specifying on neutron stars, this threshold depends on the mass and equation of state. The presence of a companion will further influence the required coupling strength for scalarization, and thus, a stable hair can be installed at a lower magnitude of coupling for those neutron stars as members of binaries. Focusing on binary neutron star mergers, we investigate this latter dynamically-driven scalarization, and find that the reduction in the threshold coupling strength seems to be more profound for symmetric binaries, while the threshold is only marginally reduced for rather asymmetric binaries. The associated scalar radiation is also discussed.
		We discover in addition a universal relation between the critical coupling strength and the stellar compactness for isolated neutron stars and perform a detailed comparison with the dynamical scalarization threshold. In synergy with such relations, one can, at least in principle, constrain the theory parameters regardless of the uncertainty in the equation of state. 
	\end{abstract}
	
	\maketitle
	
	\section{Introduction}\label{secI}
	
	Pulsar binary timing observations and gravitational waves are among the most stringent tests that have ever been implemented to scrutinize the strong field regime of Einstein's theory (e.g., \cite{frei12,abbo17,abbo19,zhao22}; see also \cite{will14,bert15,shao17} for reviews on this topic). Although general relativity (GR) gains great success in these two tests, some fundamental physical aspects, such as its non-renormalizability, are difficult to be addressed when pursuing a unification with quantum theories. An extension to GR thus warrants further investigation. Among a variety of possibilities, including a single scalar field to a gravitational theory may be the simplest modification. Such theories can be constructed by coupling a scalar field to curvature invariants of different orders as suggested by the attempts to quantize gravity, e.g., the Ricci scalar and self-contraction of the Riemann tensor. However, if we restrict ourselves to the theories admitting 2nd order field equations, the landscape of the possible candidates is narrowed down significantly, and can be parameterized as Horndeski action with scalar field coupling functions to be assumed \cite{koba19}. 
	
	We here focus on a subset of the Horndeski theories, known as scalar-Gauss-Bonnet theory (SGB). A specific property of scalarized NSs in such theories is that they have less (gravitational) mass than their GR counterpart for the same central energy density \cite{pani11,done18}. It may be worth mentioning that this is in sharp contrast to the classical scalar-tensor theory, where the scalar field tends to increase the maximum allowed NS masses. 
	In addition, as opposed to scalar-tensor theory, black holes (BHs) in SGB can circumvent the no-scalar-hair theorems and be imbued with a scalar field \cite{kant96,tori97}. Depending on the exact form of the coupling function between the scalar field and the Gauss-Bonnet invariant, either the GR compact objects can be solutions to the SGB field equations with a zero scalar field \cite{done18bh,silv18,anto18,cunh19,east21}, or the compact objects should always be endowed with scalar hair in the shift-symmetric SGB gravity \cite{kant96,tori97,pani09,soti14,benk16,ripl20a}. In the present paper, we focus on the former case, which admits spontaneous scalarization that is a nonlinear development of the scalar field once a certain threshold of the coupling strength is exceeded \cite{damo93}.
	
	The formalism of SGB is rather complicated compared to those of scalar-tensor theories for example. Thus the so-called \emph{decoupling} limit, i.e., neglecting back-reaction of the scalar field on Einstein's and Euler's equations, is often adopted in the first studies of nonlinear dynamics of compact objects \cite{benk16,silv21}. As a matter of fact, this approximation captures well both qualitatively and quantitatively the scalar field dynamics for realistic magnitudes of the scalar field \cite{ripl20a,kuan21,done21}. Furthermore, the bifurcation point of scalarization, where the GR black hole and NS loses stability and acquires scalar hair, can be determined without approximation in the decoupling limit \cite{done22}. Investigating the threshold of scalarization for coalescing binary NSs is one of our goals. 
 
    The richness/complexity of SGB does not stop at this level provided that different coupling functions may give rise to completely distinct physics. As commented above, there are no non-scalarized NSs and BHs in the shift-symmetric SGB, where the coupling function reads \cite{kant96,tori97,pani09},
	\begin{align}
		f(\varphi)=\varphi.
	\end{align}
   A spontaneous activation of the scalar field cannot be witnessed for the above coupling function therefore. However, one can observe such spontaneous scalarization once the near-horizon spacetime curvature exceeds a certain threshold for a coupling function having the form $f(\varphi)\sim\varphi^2 + \mathcal{O}(\varphi^3)$. Although keeping only the lowest order square term renders instabilities \cite{blaz18,mina19,mace19}, a convenient and well-behaved choice, having the same leading order expansion with respect to the scalar field, is the following one \cite{danc21}:
	\begin{align}\label{eq:cpling}
		f(\varphi)=\frac{\iota}{2\beta}\Big[1-e^{-\beta\varphi^2}\Big],
	\end{align}
	where $\beta$ is a dimensionless parameter, $\iota=\pm1$ and thus two flavors of the coupling are contained.
	For $\iota=1$ the no-scalar-hair theorems are circumvented and scalarized states exist for both NSs and BHs \cite{done18bh,silv18,done18}. On the other hand, scalarization can only manifest in NSs when $\iota=-1$ \cite{done18}, i.e., no-hair theorem applies to BHs (for an exception  in the case of rotating BHs see \cite{dima20,done20,herd21,bert21}).
	
	Dynamics of spontaneous scalarization in SGB gravity was investigated both for isolated BHs \cite{ripl20b,done21} and isolated NS \cite{silv18,kuan21}. When it comes to NSs in coalescing binaries, though, the collective effects on scalarization alter the picture \cite{bara13,shib14,card20} and the dynamical scalarization can be observed. More precisely, the dynamical development of the scalar hair can be activated as the two inspiralling compact objects approach each other even when one or both of the individual objects do not have the critical compactness to be scalarized when isolated. We should mention that for BHs, the process is usually the other way around and scalarized BHs can lose their scalar field as they merge \cite{silv21} (for an exception and scalar field growth during merger related to the spin-induced scalarization see \cite{Elley:2022ept}).
 
	This scenario can realize only for a coupling strength larger than a threshold, which is the subject of the present article, and when the distance between the objects is small enough, i.e., shortly before the merger of BHs or NSs \cite{bara13,shib14}. Up to now, the effort in exploring the dynamical scalarization phenomenon in SGB gravity was directed mainly towards binary BHs \cite{silv21,east21,ares22,done21L} while the NS mergers remain largely unexplored. Coalescing binary NSs is studied only recently for the specific coupling function $f(\varphi)\propto \varphi$ \cite{east22}(i.e., the shift-symmetric SGB). However, the coupling function they adopted for this first study of  binary NSs in SGB does not admit the phenomenon of scalarization. 
	
	In the present work, we aim to numerically demonstrate scalarization in binary NSs, while we work in the decoupling limit as the first step. A special emphasise is put on the condition admitting scalarization with dynamical origin. In particular, we study the threshold on the coupling strength of scalar field to the Gauss-Bonnet invariant such that the scalarization can occur in the pre-merger stage, which is shown to be different from the threshold applying to isolated NSs. The mismatch of the critical magnitude of the coupling then indicates that the scalarization in coalescing binaries is triggered differently than that in a single star; similar shift in the critical coupling strength is observed in scalar-tensor theories (e.g., \cite{shib14,card20}; see also a recent review \cite{done22}). 
	
	The article is structured as follows: We recap the theory in Sec.~\ref{setup}, focusing on the phenomenon of spontaneous scalarization and speculating possible constraints that may be placed on the theory parameters with a use of the novelly established universal relation. 
	We then turn to consider dynamical scalarization occurring in coalescing binaries in Sec.~\ref{BNS}, where a reduction in the threshold coupling strength for endowing scalar hair to NSs is elucidated. In this section, we also compute the scalar radiation associated with dynamical scalarization; accordingly, some possible implication of GW170817 are commented. We finally offer a conclusion and discussion in Sec.~\ref{discuss}. 
	The indices for the spacetime coordinates are denoted by Greek letters, while Latin ones are for indices for space coordinates. In addition, we work in the unit $c=1=G$.

	\section{scalar Gauss-Bonnet theory}\label{setup}
	
	The action in the SGB theory we consider is 
	\begin{align}\label{eq:quadratic}
		S=&\frac{1}{16\pi}\int d^4x \sqrt{-g} 
		\left[R - 2\nabla_\mu \varphi \nabla^\mu \varphi + \lambda^2 f(\varphi){\cal R}^2_{\rm GB} \right] \nonumber\\
		&  + S_{\rm matter}  (g_{\mu\nu},\Psi_m),
	\end{align}
	for which the associated field equations are summarised as
	\begin{align}\label{eq:einstein}
		R_{\mu\nu}- \frac{1}{2}R g_{\mu\nu}=&\,\, 2\nabla_\mu\varphi\nabla_\nu\varphi -  g_{\mu\nu} \nabla_\alpha\varphi \nabla^\alpha\varphi \nonumber\\
		&+ 8\pi T_{\mu\nu} - \lambda^2 \Gamma_{\mu\nu},
	\end{align}
	and
	\begin{align}\label{eq:KG}
		\nabla_\mu\nabla^\mu\varphi= -  \frac{\lambda^2}{4} \frac{df(\varphi)}{d\varphi} {\cal R}^2_{\rm GB}.
	\end{align}
 Here, $R$ and $R_{\mu\nu}$ denote the Ricci scalar and Ricci tensor, respectively, and $\Psi_m$ collectively denotes the matter fields, whose matter energy momentum tensor is $T_{\mu\nu}$. In addition, one can tell from the action that $\lambda$ has the same dimension as length.
	As introduced in Sec.~\ref{secI}, the scalar field, $\varphi$, couples to the Gauss-Bonnet invariant,
	\begin{align}
		{\cal R}^2_{\rm GB}=R^2 - 4 R_{\mu\nu} R^{\mu\nu} + R_{\mu\nu\eta\sigma}R^{\mu\nu\eta\sigma},
	\end{align}
	through a function $f(\varphi)$, where $R_{\mu\nu\eta\sigma}$ denotes the Riemann tensor. It is worth mentioning that ${\cal R}^2_{\rm GB}$ can be negative for NSs although it is always positive for BHs.
	On the right hand side of Eq.~\eqref{eq:einstein}, the back-reaction of the scalar field induced from ${\cal R}^2_{GB}$ is encoded in $\Gamma_{\mu\nu}$, defined as
	\begin{align}
		\Gamma_{\mu\nu}=& - R(\nabla_\mu\Psi_{\nu} + \nabla_\nu\Psi_{\mu} ) - 4\nabla^\alpha\Psi_{\alpha}\left(R_{\mu\nu} - \frac{1}{2}R g_{\mu\nu}\right) \nonumber\\
		& + 4R_{\mu\sigma}\nabla^\sigma\Psi_{\nu} + 4R_{\nu\sigma}\nabla^\sigma\Psi_{\mu} \nonumber\\
		&- 4 g_{\mu\nu} R^{\alpha\beta}\nabla_\alpha\Psi_{\beta} 
		+ 4 R^{\alpha}_{\;\mu\beta\nu}\nabla^\beta\Psi_{\alpha}, 
	\end{align}  
	with $\Psi_{\mu}= \lambda^2 \displaystyle{\frac{df(\varphi)}{d\varphi}}\nabla_\mu\varphi$.

	\subsection{Scalarization of isolated NSs}\label{sponscal}
	
	In the present paper, we focus on scalar field in NSs and adopt the coupling function \eqref{eq:cpling}, for which a vanishing $\varphi$ naturally satisfies Eq.~\eqref{eq:KG},  and thus GR solutions are always solutions to the considered theory. However, the perturbation of interior scalar field can develop to scalarize the star via tachyonic instability depending on the theory parameters and the nuclear matter equation of state (EOS). In particular, there is a critical coupling strength $\lambda_{\rm bif}$ for some specific stellar mass ($M_\star$) and EOS beyond which the NS is susceptible to scalarization; for example, NSs with mass $\gtrsim1.4\,\,M_{\odot}$ and EOS APR4 will be imbued with a stable scalar hair for a coupling strength $\lambda\ge13\,M_\odot$ when $\iota=-1$. It may be worth mentioning that the condition allowing for scalarization is solely determined by $\lambda$, while the quantitative profile of the scalar field is determined jointly by $\beta$, $\lambda$, and $\iota$  \cite{done18} (see also below). As seen later, the threshold of $\lambda$ to allow for a scalarized NS with a certain mass will however change in the presence of a binary companion.

	For isolated and non-spinning NSs, the metric of spherically symmetric NSs can be expressed as
	\begin{align}\label{eq:metric_sph}
		ds^2=-\alpha(r)^2dt^2+X(r)^2dr^2+r^2(d\theta^2+\sin^2\theta d\phi^2)
	\end{align}
	with functions $\alpha$ and $X$ depending solely on the radius $r$.
	Owing to the respected symmetry, perturbations of the scalar field can be decomposed into harmonic components in terms of the spherical harmonics function $Y_{\ell m}$, labeled by the eigen-values $\{\ell,m\}$, as
	\begin{align}
		\varphi=\frac{u(r)}{r}e^{-i\omega t}Y_{\ell m}(\theta,\phi).
	\end{align}
	The radial part, $u(r)$, is governed by the linearised equation (see Eqs.~(2.18) and (2.19) of \cite{done18}), 
	\begin{align}
		\frac{\alpha}{X}\frac{d}{dr}\left[\frac{\alpha}{X}\frac{du(r)}{dr}\right] + [\omega^2-U(r)]u(r)=0
	\end{align}
	with the effective potential $U(r)$ given by
	\begin{align}\label{eq:Upotential}
		U(r)=&\alpha^2\Bigg[ \frac{X^{-2}}{r}\left(\frac{d\ln\alpha}{dr}-\frac{d\ln X}{dr}\right)+\frac{\ell(\ell+1)}{r^2} \nonumber\\
		&-\frac{\lambda^2\iota}{4}{\cal R}_{\rm GB}^2 \Bigg],
	\end{align}
	and the frequency of this specific mode $\omega$.
	
	Specifying fluid to be perfect, the energy-stress tensor is given by,
	\begin{align}
		T_{\mu\nu}=\rho hu_\mu u_\nu + Pg_{\mu\nu},
	\end{align}
	where $\rho$ is the rest-mass density, $u^\mu$ is the four-velocity, and \text{$h=(\epsilon+P)/\rho$} is the specific enthalpy for the energy density $\epsilon$ and the pressure $P$. The Gauss-Bonnet invariant for non-scalarized NSs (i.e., $R_{\mu\nu}-R g_{\mu\nu}/2=8\pi T_{\mu\nu}$ is obeyed) then has the form (see Eq.~(2.20) of \cite{done18})
	\begin{align}
		{\cal R}_{\rm GB}^2 = \frac{48 m^2}{r^6} - \frac{128}{r^3}(2\pi P r^3+m)\epsilon,
	\end{align} 
	where the mass $m$ is defined by
	\begin{align}
		m=\frac{r}{2}(1-X^{-2}).
	\end{align}
	It can be seen that ${\cal R}_{\rm GB}^2$ is negative at center of NSs, and gradually grows to positive values as approaching stellar surface at which it reaches the peak and then it slowly drops to zero toward the infinity.
	The tachyonic instability is present in a non-scalarized NS if the effective potential $U(r)$ suggests a negative eigen-value of $\omega^2$ for a harmonic mode. In order for that to occur, the positive part of $\lambda^2\iota{\cal R}^2_{\rm GB}$ should be large enough somewhere inside the star so that the potential can be sufficiently `deep' to harbor a scalarized state. 
	
	In addition, the absence of $\beta$ in $U(r)$ indicates that whether an isolated NS can be stably endowed with a scalar field is independent of $\beta$ \cite{done18}. We denote as $^{\pm}\lambda_{\rm bif}$ the critical coupling strength to grow a scalar field for $\iota=\pm1$. It should be noted that, as a matter of fact, more than one scalarized solution exist and the different solutions are labeled by the number of scalar field nodes. Clearly, the threshold $^{\pm}\lambda_{\rm bif}$ varies for classes of solutions with different number of scalar field nodes. Nonetheless, we focus on the nodeless scalar field solutions since they are the only ones that are stable and can realize in some dynamical scenario \cite{blaz18,kuan21}. 
	
	The configuration of the nodeless scalar field in `hairy' NSs is quantitatively different for the two flavors of the coupling function \eqref{eq:cpling}. In particular, the maximum of $\varphi$ locates at the stellar center for $\iota=-1$ (see Fig.~3 of \cite{kuan21}), while the peak of $\varphi$ lies at a finite distance close to the star's surface when $\iota=1$ (see Fig.~3 in the supplemental material of \cite{kuan21}). The difference roots in the source term of the scalar field. In particular, $\iota{\cal R}_{\rm GB}^2$ has the maximum nearby the stellar surface for $\iota=1$, where the scalar field ``condensates''. On the other hand, the maximum of $\iota{\cal R}_{\rm GB}^2$ locates at the center if $\iota=-1$, which then suggests a profile of $\varphi$ that peaks at the center.

	\begin{figure}
		\centering
		\includegraphics[width=\columnwidth]{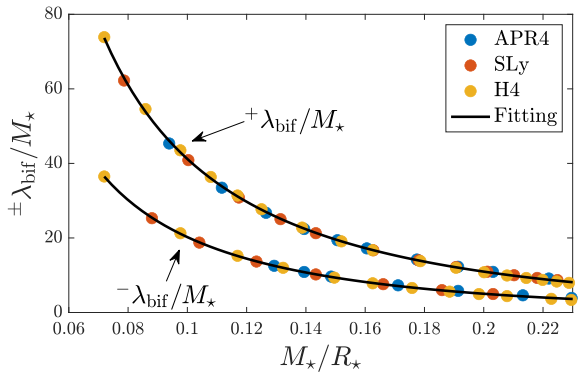}
		\caption{Universal relation between the stellar compactness and the mass-scaled critical coupling strength. The filled circles show the values allowing a marginal scalarization in the star with indicated mass. Three EOSs are considered in the plot, viz.~APR4 (blue), SLy (red), H4 (yellow). The overlapped solid lines represent the fitting formulae \eqref{eq:fit_m} and \eqref{eq:fit_p}, serving as a boundary separating the parameter space admitting scalarized states (above the lines) from that prohibits spontaneous scalarization (below the lines).
		} 
		\label{fig:uni}
	\end{figure}
	
	For nodeless scalar hairs, we find for the first time and plot in Fig.~\ref{fig:uni} the universal relations between the stellar compactness\footnote{Just to eliminate possible confusion, a NS with $M_\star=1.4\,\,M_\odot$ and $R_\star=12$~km has a compactness of ${\cal C}=0.172$.}, defined as ${\cal C}=M_\star/R_\star$ with $R_\star$ being the radius of the star, and the threshold coupling strength $^{\pm}\lambda_{\rm bif}$ for three EOSs, viz.~APR4, SLy, and H4. Two of the considered EOSs, namely APR4 and SLy, have similar stiffness, while the H4 one is the stiffest among the three. Each of the EOSs belongs to different groups proposed in \cite{kuan22} (see Tab.~1 therein), categorized by some internal oscillation properties. The fitting formulae, derived from the data of $0.07 \alt {\cal C} \alt 0.24$,  are expressed as 
	\begin{subequations}
		\begin{align}
			\frac{^{-}\lambda_{\rm bif}/M_\star}{30} \simeq \frac{0.819{\cal C}^2-0.586{\cal C}+0.127}{{\cal C}-0.021}, \label{eq:fit_m} \\
			\intertext{and}
			\frac{^{+}\lambda_{\rm bif}/M_\star}{30} \simeq \frac{1.422{\cal C}^2-1.040{\cal C}+0.247}{{\cal C}-0.023}. \label{eq:fit_p}
		\end{align}
	\end{subequations}
	The denominator of both equations should theoretically be the compactness without an abstraction of a small but non-zero number. This latter non-zero number attributes to the limited datum used for fitting; in particular, we consider finitely many models in a specific range, which is large enough to account for any possible NS that could manifest in nature but not complete in theoretic aspect.

	\subsection{Possible EOS-insensitive constraints on SGB gravity with a massless scalar field}\label{obs}
	The relations \eqref{eq:fit_m} and \eqref{eq:fit_p} may be used to set constraints on $\lambda$ that are immune to the uncertainty of nuclear EOS. Here we will focus on the constraints coming from the orbital period decrease of pulsar binaries that are the most stringent ones \cite{shao17,zhao22}. In particular, for those close binaries involving pulsar(s) where the mass and radius of the NS can be estimated with relatively good accuracy, a non-detection of scalar dipole radiation could place an \emph{EOS-insensitive} bound on the coupling parameter $^{\pm}\lambda$ for both $\iota=\pm1$ cases.
    For example, PSR J0740+6620 are estimated from its X-ray spectral-timing to have a mass of $M_\star\simeq 2.01$--$2.15\,\,M_\odot$ and an equatorial radius $R_\star\simeq11.41$--$16.3$~km \cite{rile21,mill21,fons21}. Running over the mass and radius ranges, we find that the most conservative upper bounds that could be inferred by a future null observation of a dipolar scalar radiation from this pulsar system, if confirmed at all, are $^{-}\lambda\lesssim 14.54\,M_\odot$ and $^{+}\lambda\lesssim 30.88\,M_\odot$ (though see below for possible relaxation by a scalar mass).
	
	In this respect, some other examples may be accreting millisecond X-ray pulsars (AMXPs) in low-mass X-ray binaries (LMXBs), such as XTE J1751-305, IGR J00291+5934 and SAX J1808.4-3658. The orbital evolution of AMXPs is much more complicated than that of the secular pulsar binaries like PSR J0740+6620 (e.g., \cite{di08,hart08}) because of the active communication between the two orbiting objects as well as the interaction between individual stars and the orbit (see a recent review \cite{di20}). Although subject to the uncertainty of the mass-transfer physics, AMXPs still have potential to set bound on the strength of scalar radiation in the future.
	
 Among the aforementioned AMXPs, the simultaneous measurement of mass and radius are arguably obtainable for XTE J1751-305 considering that the stellar spin is saturated by $r$-mode instability \cite{ande14}, where $M_\star=1.59$--$1.91\,M_\odot$ and $R_\star=11.8\pm0.9$~km are inferred. Together with the fitting formulae \eqref{eq:fit_m} and \eqref{eq:fit_p}, the most conservative bounds $^{-}\lambda\lesssim 11.17\,M_\odot$ and $^{+}\lambda\lesssim 23.77\,M_\odot$ could be placed if the null detection of scalar radiation could be acclaimed with future observations. The masses and radii of the other two AMXPs are however not so well limited (see, e.g., \cite{bhat01,leah08}). On the other hand, by scrutinising the spectroscopic data on multiple thermonuclear bursts, the X-ray-loud NS in the LMXB 4U 1820-30 are estimated to have $M_\star=1.58\pm0.06\;M_\odot$ and $R_\star=9.1\pm0.4$~km \cite{guver10}. This latter system has a potential to limit the coupling strength to be $^{-}\lambda\lesssim 6.35\,M_\odot$ and $^{+}\lambda\lesssim 14.22\,M_\odot$ positing again that the scalar dipolar radiation is absent in the measured orbital decay \cite{van93}. 
	
	We note, however, that the fitting formulae \eqref{eq:fit_m} and \eqref{eq:fit_p} are found for \emph{isolated} and \emph{non-rotating} NSs. The influence of spin and a companion on the threshold is expected from the results of scalar-tensor theory (see, e.g., \cite{done22}). We study the latter effect in Sec.~\ref{dynscalar}, while deferring the investigation about the spin effect to future work.
	As we will show later, $^{\pm}\lambda_{\rm bif}$ is lower for a NS if it is a member of binary instead of being reclusive since the maximum of $\iota{\cal R}_{\rm GB}^2$ increases as the binary approaches merger thus hinting at a smaller $^{\pm}\lambda_{\rm bif}$. 
	Accordingly, a more stringent bound may possibly be placed if scalar-induced phenomena are absent in binary pulsars and/or coalescing binaries. Similar reduction in the scalarization threshold is observed and elucidated in scalar-tensor theory \cite{bara13,shib14} but until now it was not studied in detail in SGB gravity neither for BHs nor for NSs. 
	In addition, developing a similar universal relation for rotating NSs will supplement in this direction, while we expect that the deviations should be small for the spins of currently observed pulsars based on the studies of other universal relations \cite{done14,papp14,chak14}. 
 
Apart from the constraints that can be potentially set by future observations in  the aforementioned manner, measuring the shrinking of the orbit of NS-white dwarf binaries is reportedly able to place a limit on $\lambda$. This was illustrated in a recent work \cite{danc21} (see also \cite{wong22}). Since there is already a good number of such observed systems with a relatively well-measured orbital decay due to gravitational wave emission, it was obtained that roughly $^{+}\lambda\lesssim 24\,M_\odot$ and $^{-}\lambda\lesssim 8\,M_\odot$. These limits are strongly dependent on the EOS, though, since they do not employ an EOS universal relations.

 \subsection{Relaxing observational constraints on SGB gravity with a massive scalar field}\label{obs}
    If the above constraints are satisfied, it will be difficult to observe any measurable deviation from GR in binary mergers. The aforementioned bounds ought to be eased, though, if the scalar field is massive since a mass of $m_\varphi$ will introduce a Yukawa suppression $e^{-r/\lambda_{\rm comp}}$ on the scalar field with a Compton length-scale $\lambda_{\rm comp}=\hbar/m_\varphi$. A tiny mass of $m_\varphi=10^{-16}$~eV is enough to evade the binary pulsar observation since the associated $\lambda_{\rm comp}\simeq2\times10^{6}$~km is much smaller than the orbital separation for the binary pulsars cited above, while still allowing for large deviations for merging NSs (e.g., \cite{rama16,yaza16}; see also below).  
    
\begin{figure}
		\centering
		\includegraphics[width=\columnwidth]{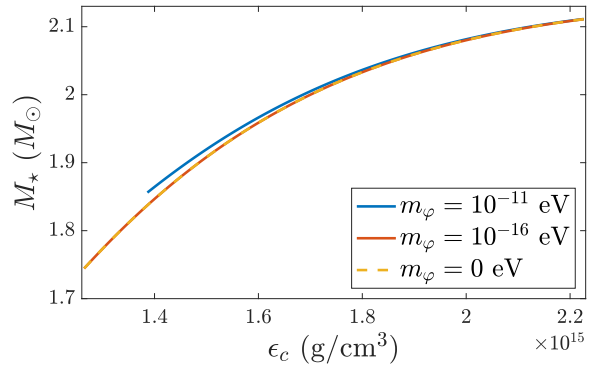}
		\caption{Stellar mass to central energy density ($M_\star-\epsilon_c$) curve for equilibrium of scalarized NSs pertaining to the EOS APR4 in the considered theory with parameters $\lambda=10\,M_\odot$, $\beta=250$ and $\iota=-1$. The blue sequence corresponds to NSs that are imbued by a scalar field with a mass of $m_\varphi=10^{-11}$~eV, while the red one represents solutions with massless scalar field. 
		} 
		\label{fig:seq}
	\end{figure}

    Interestingly, such tiny scalar field mass does not change the stellar structure in any practical sense. In Fig.~\ref{fig:seq}, we plot the mass of scalarized NSs as a function of the central energy density $\epsilon_c$ for three different masses~$m_\varphi=0$~eV, $10^{-11}$~eV, and $10^{-16}$~eV while fixing $\lambda=10M_\odot$(=14.77~km), $\beta=250$ and $\iota=-1$. Only the scalarized equilibria are plotted while the solutions with $\varphi=0$ are omitted for better visualization. As seen, the change in the NS equilibrium properties differs negligibly even for  $m_\varphi=10^{-11}$~eV while the most significant change is the point of bifurcation that moves to larger central energy densities. This gives us the confidence that the scalar field dynamics for a massive scalar field with $m_\varphi=10^{-16}$~eV, that evades binary pulsar observations, will be practically indistinguishable from the massless case which allow us to consider values of $\lambda$ larger than the above mentioned constraints as demonstrated in the next section.
	
	\subsection{Projected constraints in the massive scalar field case by a network of observations}\label{massive}
	
In the previous subsection, we discussed how even a tiny scalar field mass can suppress the scalar gravitational radiation for binary pulsars. The smaller the separation between the stars, though, the larger the scalar field mass should be in order to smear out the scalar-induced effects. Thus, the constraints one can set on massive SGB theories from the merger of compact objects are much stronger due to the reduced separation. In this subsection, we put these quantitative considerations on a more solid ground and speculate on the strongest constraints could be placed by a network of observations.
 
\begin{figure}
		\centering
		\includegraphics[width=\columnwidth]{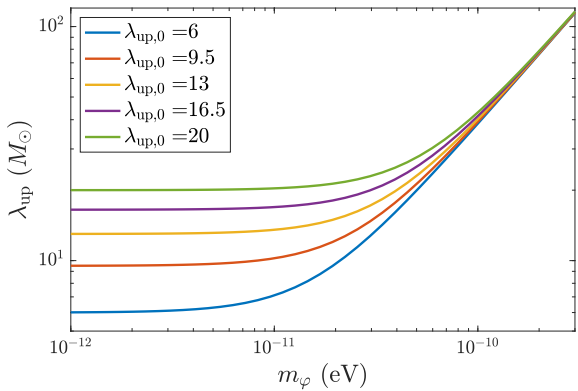}
		\caption{Upper bound on the coupling strength $\lambda$ as a function of $m_\varphi$ for several assumed values of $\lambda_{\rm up,0}$, which is the bound if the scalar field is massless. We assumed $|{\cal R}_{\rm GB}^2|_{\rm max}=10^{-3}\,M_\odot^{-4}$.
		} 
		\label{fig:lmb_mod}
	\end{figure}

	The inclusion of a scalar mass via a potential having the form $V=2m_\varphi^2\varphi^2$ in the action \eqref{eq:quadratic} renders the equation,
	\begin{align}
		\nabla_\mu\nabla^\mu \varphi=-\frac{\lambda^2}{4}\frac{df}{d\varphi}{\cal R}_{\rm GB}^2 + m_\varphi^2\varphi,
	\end{align}
	for the scalar field. For a small perturbation of $\varphi$, the right-hand side can be approximated to give
	\begin{align}\label{eq:stability_modified}
		\nabla_\mu\nabla^\mu \varphi\simeq \Big(-\frac{\iota\lambda^2}{4}{\cal R}_{\rm GB}^2+m_\varphi^2\Big) \varphi.
	\end{align}
	In addition, we find for a typical BNS that the value of ${\cal R}_{\rm GB}^2$ lies in the range of 
	\begin{align}\label{eq:GBrange}
		-1\times 10^{-2}\,M_\odot^{-4} \lesssim  {\cal R}_{\rm GB}^2 \lesssim 1\times10^{-3}\,M_\odot^{-4}
	\end{align}
	throughout the last stages of inspiral until the onset of merging. During the merger, the magnitudes of the lowest and highest values of ${\cal R}^2_{\rm GB}$ are significantly stretched by almost an order of magnitude, which will then relax to the values for the formed remnant. For this range \eqref{eq:GBrange} of values, a mass such that
	\begin{align}
		m_\varphi &>  \frac{\lambda}{2} \sqrt{|{\cal R}_{\rm GB}^2|_{\rm max}} \nonumber\\
		&\approx 3.34\times 10^{-11} \left(\frac{\lambda}{10\,M_\odot}\right)
		\left(\frac{\sqrt{|{\cal R}_{\rm GB}^2|_{\rm max}}}{0.05\,M_\odot^{-2}}\right)\,\, {\rm eV}
	\end{align}
	will then essentially quench the growth of scalar field (cf.~Fig.~\ref{fig:lmb_mod}), where $|{\cal R}_{\rm GB}^2|_{\rm max}$ denotes the maximum of $|{\cal R}_{\rm GB}^2|$. The reason is that the term in the brackets on the right-hand side of Eq.~(\ref{eq:stability_modified}) will become positive thus forbidding the existence of any bounds state of the generalized potential (\ref{eq:Upotential}). 
	
	Therefore, the upper bound for the coupling strength ($\lambda_{\rm up}$) that could be obtained from a certain network of observations, both binary pulsar and merger ones, depends on the scalar mass. It is expressed as
	\begin{align}
		\lambda_{\rm up}(m_\varphi)=\sqrt{\lambda_{\rm up,0}^2+\frac{4m_\varphi^2}{|{\cal R}_{\rm GB}^2|_{\rm max}}},
	\end{align}
	where $\lambda_{\rm up,0}=\lambda_{\rm up}(0)$ is the bound for zero scalar field mass. We note that this expression is applicable for $\iota=\pm1$. In Fig.~\ref{fig:lmb_mod}, we show how the bound in massless theory (i.e., $\lambda_{\rm up,0}$) would be modified by $m_\varphi$, where we see that the constraint set for a massive theory does not deviate in a noticeable way from what could be inferred for a massless theory with $m_\varphi<10^{-11}$~eV.

	\section{Coalescing Binary Neutron Stars}\label{BNS}
	
	Having recapped the formulae of SGB, introduced the novel universal relations, illustrated how an EOS-insensitive constraint could be placed and how a tiny scalar mass is sufficient to relax the constraints on parameters, we now proceed to the main part of the present paper, where we aim at studying binary NS coalescence in SGB gravity with an emphasis on dynamically-triggered scalarization. In this section, however, we consider only the $m_\varphi=0$ case but the dynamics of $\varphi$ and the rendered scalar radiation would be practically unchanged if the scalar mass is less than a few times of $10^{-13}$~eV since the associated Compton wavelength ($\agt10^3$~km) is longer than the binary separation and the wavelength of scalar waves in the late inspiralling phase that we consider in this paper. 
	For this purpose, the simplified metric \eqref{eq:metric_sph} for stationary and spherically-symmetric spacetime should be generalised to describe the spacetime where a coalescing binary is embedded. We express the metric hereafter as
	\begin{align}
		ds^2=-\alpha(t,\boldsymbol{x})^2dt^2+\gamma_{ij}(t,\boldsymbol{x})(dx^i + \beta^i dt)(dx^j+\beta^jdt),
	\end{align}
	where the lapse function $\alpha$, the shift vector $\beta^k$, and the induced metric on three-dimensional hypersurfaces $\gamma_{ij}$ vary with time $t$ and position $\boldsymbol{x}$. 
	
	In the decoupling limit that we adopt in the present work, we will ignore the influence of scalar field on the \text{3+1} decomposed Einstein field equations, thus leaving them identical to those in GR. As commented in the introduction, this approximation gives accurate results for the scalar field dynamics in the cases where the scalar field's magnitude is limited to low or moderate values. In addition, the  threshold for scalarization of the GR solutions can be obtained exactly within the decoupling limit, i.e., it is not influenced by this approximation. 
	
	The field equation of the scalar field in 3+1 decomposition has the form:
	\begin{align}
		(\partial_t-\beta^{k}\partial_k)\varphi =& -\alpha K_\varphi, \nonumber\\
		(\partial_t-\beta^{k}\partial_k)K_\varphi =& -\alpha D_iD^{i}\varphi-(D_i\alpha)D^{i}\varphi+\alpha K K_\varphi \nonumber\\
		&- \frac{\iota \alpha\lambda^2}{4} \varphi e^{-\beta\varphi^2}  {\cal R}^2_{GB},
	\end{align}
	where $K$ is the trace of the extrinsic curvature tensor $K_{ij}$, and $K_\varphi=-n^{a}\nabla_a\varphi$ for $n^{a}$ the unit vector normal to spatial hypersurfaces. In the second equation we have used the specific coupling \eqref{eq:cpling}. In addition, the Gauss-Bonnet invariant ${\cal R}^2_{GB}$ can be expressed as
	\begin{align}
		{\cal R}^2_{GB} =& 8E_{ij}E^{ij}-4\Omega_{ijk}\Omega^{ijk} -\frac{256\pi^2}{3}(\rho h-P)(\rho h+2P),
	\end{align}
	where the tensors $E_{ij}$ and $\Omega_{ijk}$, defined by
	\begin{align}
		E_{ij} =& ^3R_{ij} + KK_{ij} - K_i^{~l}K_{lj} \nonumber\\
		&- \frac{16\pi}{3}\rho_h\gamma_{ij}-4\pi\left( S_{ij}-\frac{1}{3}\gamma_{ij}S_k^{~k} \right),
	\end{align}
	and
	\begin{align}
		\Omega_{klj}=D_k K_{lj}-D_lK_{kj}-4\pi(\gamma_{jk}J_l-\gamma_{jl}J_k),
	\end{align}
	characterise the electric and magnetic components of the Weyl tensor. 
	Here $^3R_{ij}$ is the Ricci tensor with respect to $\gamma_{ij}$, whose determinant is denoted as $\gamma$, and we have introduced 
	\begin{align}
	    \rho_{\rm h}=T_{\mu\nu}n^{\mu}n^{\nu},
	\end{align}
	\begin{align}
		J_k=-T_{\mu\nu}n^\mu\gamma^\nu{}_k=\rho hwu_k,
	\end{align}
	and
	\begin{align}
		S_{ij}=T_{\mu\nu}\gamma^\mu{}_i\gamma^\nu{}_j=\rho h u_iu_j+P\gamma_{ij}
	\end{align}
	with the Lorentz factor $w=-u_\mu n^\mu$.
	
	Defining the trace-free extrinsic curvature by
	\begin{align}
		A_{ij} = K_{ij} - \frac{1}{3}\gamma_{ij} K,
	\end{align}
	and its conformally-related tensor $\tilde{A}_{ij}=W^2A_{ij}$ for $W=\gamma^{-1/6}$, we can rewrite the tensors $E_{ij}$ and $\Omega_{ijk}$ as,
	\begin{align}
		E_{ij} =& ^3R_{ij} -4\pi\left(S_{ij}-\frac{1}{3}\tilde{\gamma}_{ij}S_{kl}\tilde{\gamma}^{kl}\right)
		\nonumber\\
		&+ W^{-2} \bigg[ - \frac{16\pi}{3}\rho_h \tilde{\gamma}_{ij}+\frac{K}{3}\tilde{A}_{ij}+\frac{2K^2}{9}\tilde{\gamma}_{ij}-\tilde{A}_i^{~l}\tilde{A}_{lj}
		\bigg],
	\end{align}
	and 
	\begin{align}
		\Omega_{klj}=&-2W^{-3}\left(\tilde{A}_{lj}\partial_k W-\tilde{A}_{kj}\partial_l W\right) 
		\nonumber\\
		&+ W^{-2}\bigg[\partial_k\tilde{A}_{lj}-\partial_l\tilde{A}_{kj}+\Gamma^{i}_{~lj}\tilde{A}_{ik}-\Gamma^{i}_{~kj}\tilde{A}_{il}
		\nonumber\\
		&+\frac{1}{3}\left(\tilde{\gamma}_{lj}\partial_kK-\tilde{\gamma}_{kj}\partial_lK\right)
		-4\pi\left(\tilde{\gamma}_{jk}J_l-\tilde{\gamma}_{jl}J_k\right)
		\bigg],
	\end{align}
	respectivley, where $\Gamma^{i}_{~jk}$ are the Christoffel symbols with respect to $\gamma_{ij}$.

	\subsection{Numerical method}
	
	\begin{figure}
		\centering
		\includegraphics[width=\columnwidth]{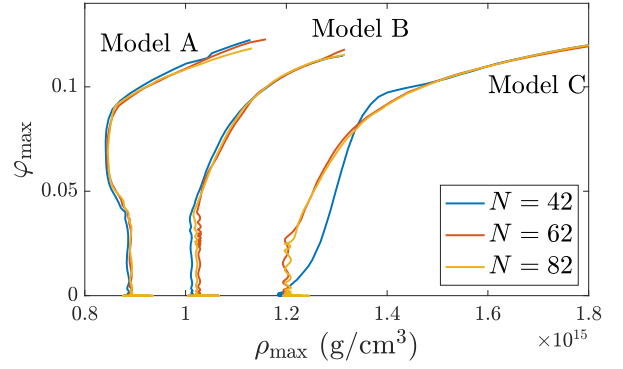}
		\caption{Maximum values of the scalar field as a function of the maximum rest-mass density ($\varphi_{\rm max}$-$\rho_{\rm max}$) from late inspiral stages to shortly after merger. Models A, B, and C in Tab.~\ref{tab:binaries} are simulated with three grid resolutions with $N=42, 62$, and $82$, respectively. The initial separation for the binaries is fixed to be 41.35\,km while the associated initial orbital frequencies are listed in the last column of Tab.~\ref{tab:binaries}. The coupling strength is chosen as $\lambda=17.42$~km, $15.50$~km, and $12.67$~km for models A, B, and C, respectively. The rest of theory parameters are set as $\beta=200$ and $\iota=-1$.
		}
		\label{fig:conv}
	\end{figure}
	
	Here we briefly summarise the technical setting for evolving binary NSs in the code \texttt{SACRA-MPI}, while we refer the interested reader to the original articles \cite{yama08,kiuc17} for more details.
	An adaptive mesh-refinement (AMR) algorithm with 2:1 refinements is implemented to construct a graded grid. For the simulations in the present article, the grid structure is set under the mirror symmetry about the orbital plane, and consists of 6 non-moving, concentric boxes for the whole binary. Within the finest fixed box, two piles of moving boxes exist for each NS with each graded in 4-levels.
	Adopting the Cartesian coordinate, one box contains $(2N+1)\times(2N+1)\times (N+1)$ points covering the domain along the $x$-, $y$-, and $z$-axis, respectively. The resolution is set by $N$, and the size of the finest box, which is a moving box, is chosen to have the dimension of 25.84 km$\times$25.84 km$\times$12.92 km.
 
 The initial datum for the simulations in this work are generated by an open source code FUKA \cite{gran10,pape21}, which is designed for GR. Since we are adopting the decoupling limit approximation, the initial data are exact for the metric and the fluid variables while the scalar field initial data is in the form of a tiny perturbation. Although both NSs are always non-scalarized at the beginning of our simulations, this inaccuracy is ``corrected'' relatively fast and the tachyonically unstable NS develops scalar field exponentially over a certain timescale (typically $\lesssim5$~ms) after the beginning of simulation.
	Not starting with consistent scalar field initial data might be unsatisfactory for accurate waveform generation \cite{corm22} but is sufficient for studying the onset of dynamical scalarization \cite{silv21,east22}. 
	
	As the last remark of this section, we confirm the convergence of each model listed in Tab.~\ref{tab:binaries}; these models will be described in detail in the following section. By implementing three different grid resolutions for the simulations, we plot in Fig.~\ref{fig:conv} the maximum values of the scalar field ($\varphi_{\rm max}$) as a function of the maximum rest-mass density of the numerical domain ($\rho_{\rm max}$), which is the central density of the heavier progenitors before merger and roughly that of the merger remnant.
	We observe a good convergence for the considered models, and we stick to $N = 62$ as the standard resolution in this article.

	\begin{table*}
		\centering
		\caption{Source parameters relevant here for the GW170817-like binaries characterised by a fixed ${\cal M}=1.186M_\odot$ with both NSs each having a mass in the range estimated by GW170817. The APR4 EOS is assumed for all the involved NSs, and the theory takes the flavor of $\iota=-1$. We performed a handful of simulations for every considered binaries by using several values of $\lambda$ and two initial separations (the associated initial mass-scaled orbital frequencies are presented as number pairs in the 6th column). For each separation, the used coupling strengths are linearly- and logarithmically-even-sampled by 21 values in the two ranges listed in the last column, respectively.
		}
		\begin{ruledtabular}
			\begin{tabular}{ccccccc}
				Model& $m_1$ $(M_\odot)$ & $m_2$ $(M_\odot)$ & $^{-}\lambda_{\rm bif,1}/m_1$ & $^{-}\lambda_{\rm bif,2}/m_2$ & $\Omega_{\rm ini}$($m_1+m_2$) & $\lambda$ for simulations ($M_\odot$) \\
				\hline
				A & 1.365 & 1.360 & 9.890 & 9.967 & (0.0270, 0.0224) &  [11.5,15] $\&$ [11.68,12] \\ 
			    \hline
				B & 1.625 & 1.149 & 6.768 & 13.950 & (0.0276, 0.0229) & [10.4,10.6] $\&$ [10.4,13] \\
				\hline
				C & 1.890 & 1.002 & 4.598 & 18.098 & (0.0292, 0.0243) & [8.56,8.8] $\&$ [8.56,11.5]
			\end{tabular}
		\end{ruledtabular}
		\label{tab:binaries}
	\end{table*}
	
	\subsection{Dynamical scalarization}\label{dynscalar}
	
	While the bifurcation point on the GR sequence of isolated NS equilibria is determined solely by the parameter $\lambda$ for a given EOS, the ratio between the masses of the stars in the binary, $q=m_2/m_1$, will influence the onset of dynamical scalarization; stipulating two binaries with the same total mass but two different mass ratios, the binary harbouring the more massive NS will undergo dynamical scalarization earlier owing to its lower value of $\lambda_{\rm bif}$.
	In general, the threshold for dynamical scalarization depends on the total mass, mass ratio, theory parameters (viz. $\lambda$, $\beta$, and $\iota$), and the EOS. Given that $\lambda_{\rm bif}/M_\star$ can be associated with stellar compactness for an isolated NS in an EOS-independent way (see Fig.~\ref{fig:uni}), we specify ourselves hereafter on the EOS APR4, and opt to quantify the effect of mass ratio when the chirp mass,
	\begin{align}
		{\cal M}=\frac{(m_1m_2)^{3/5}}{(m_1+m_2)^{1/5}},
	\end{align}
	is fixed. A motivation for this choice comes from the fact that ${\cal M}$  can be estimated rather accurately from gravitational waveforms (typically within an error of order $10^{-3}\%$; see, e.g., \cite{cutl94}). We consider three binaries with the chirp mass of the gravitational wave event GW170817 (i.e., ${\cal M}=1.186\,M_\odot$). Their properties are collated in Tab.~\ref{tab:binaries}, including the individual masses of the NS members and the respective bifurcation threshold of $^{-}\lambda$ of each of the binary component, viz. $^{-}\lambda_{\rm bif,1}$ and $^{-}\lambda_{\rm bif,2}$. We consider only $\iota=-1$ for studying dynamical scalarization in the present article. The results for $\iota=1$, however, seem to suggest otherwise, and will be studied in a separate work.

	\begin{figure}
		\centering
		\includegraphics[width=\columnwidth]{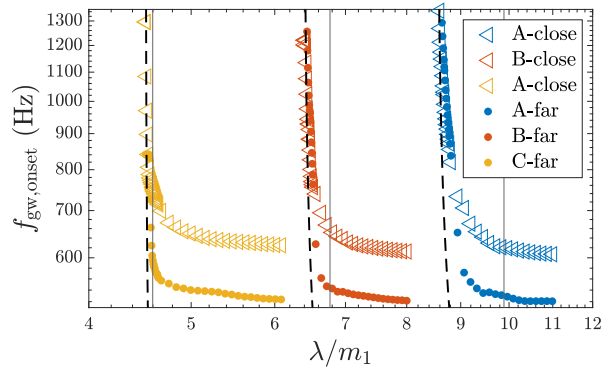}
		\caption{Relation between the coupling strength and the gravitational wave frequency at the occurrence of scalarization, which is defined as the moment when the maximum of $\varphi$ exceeds $10^{-4}$. The black vertical lines show the mass-scaled threshold for spontaneous scalarization of the heavier NS member for each binaries listed as $^{-}\lambda_{\rm bif,1}/m_1$ in Tab.~\ref{tab:binaries}. Each simulation is labeled by the model names listed in Tab.~\ref{tab:binaries} with `close' denoting the larger $\Omega_{\rm ini}$ case and `far' otherwise. The dashed lines show the tendency of the reduction of critical threshold, predicted by the Keplerian orbit, as a leading-order estimate.
		} 
		\label{fig:bndry}
	\end{figure}
	
    Identifying the occurrence of scalarization as the moment when the magnitude of the scalar field exceeds $10^{-4}$, in Fig.~\ref{fig:bndry} we show the gravitational wave frequency at the onset of scalarization as a function of the dimensionless coupling strength $^{-}\lambda/m_1$. We consider two initial separations for each model so as to reinforce the dynamical origin of scalarization. The associated initial orbital angular frequencies $\Omega_{\rm ini}$ are collated in the second last column of Tab.~\ref{tab:binaries}, where the two numbers for a model correspond to two separations set to the initial data. 
    The onset of dynamical scalarization should not depend on the chosen initial separation, while the spontaneous scalarization kicks in at the beginning of the simulations regardless initial separation and completes over a dynamical timescale. 
    In particular, the heavier (less massive) NS in each model is spontaneously scalarized if the dimensionless coupling strength that we employ in a particular simulation is greater than the threshold $^{-}\lambda_{\rm bif,1}/m_1$ ($^{-}\lambda_{\rm bif,2}/m_2$) listed in the fourth (fifth) column of Tab.~\ref{tab:binaries}.
    
    These thresholds corresponding to $^{-}\lambda_{\rm bif,1}/m_1$ for each binary are plotted by the vertical black lines in Fig.~\ref{fig:bndry}. On the right with respect to this line, the massive NS companion should always be scalarized irrespective of whether it is isolated or in binary. On the other hand, this is not the case for the region of small $\lambda/m_1$ to the left, where only dynamical scalarization can endow the NSs with a scalar field\footnote{Clearly, since the less massive companion has $^{-}\lambda_{\rm bif,2}/m_2 > ^{-}\lambda_{\rm bif,1}/m_1$, it is also for sure non-scalarized to the left of the vertical line.}.   Indeed, this is the phenomenon we observe: the scalarization kinks in at smaller binary separation, thus larger gravitational-wave frequencies, for decreasing $\lambda/m_1$.  Although we plot in Fig.~\ref{fig:bndry} only the threshold of dynamical scalarization in the primary since it is easier to be installed a stable hair, we do witness a mutual scalarization for model A, where the secondary has a more or less the same mass as the primary.
	
	\begin{figure}
		\centering
		\includegraphics[width=\columnwidth]{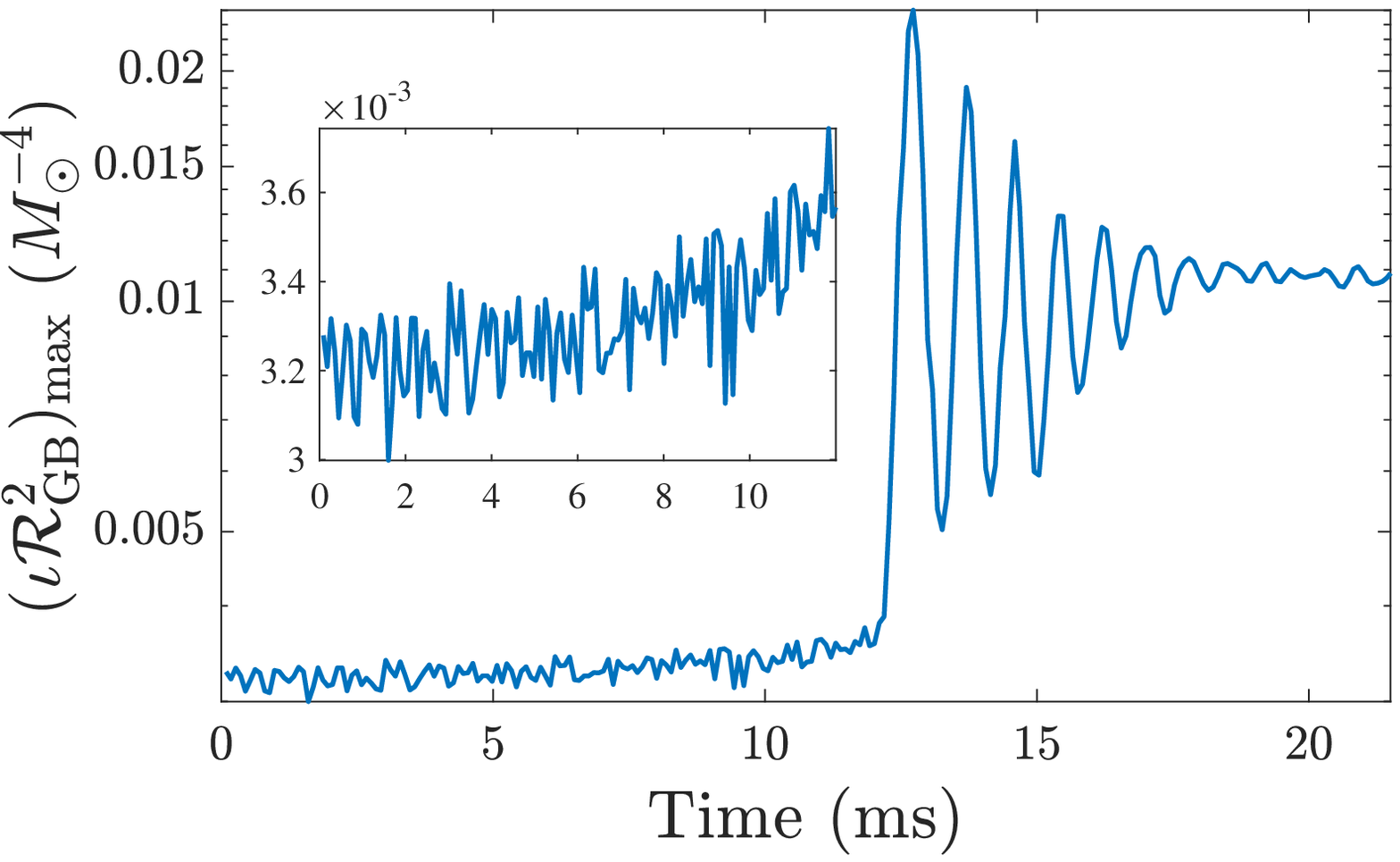}
		\includegraphics[width=\columnwidth]{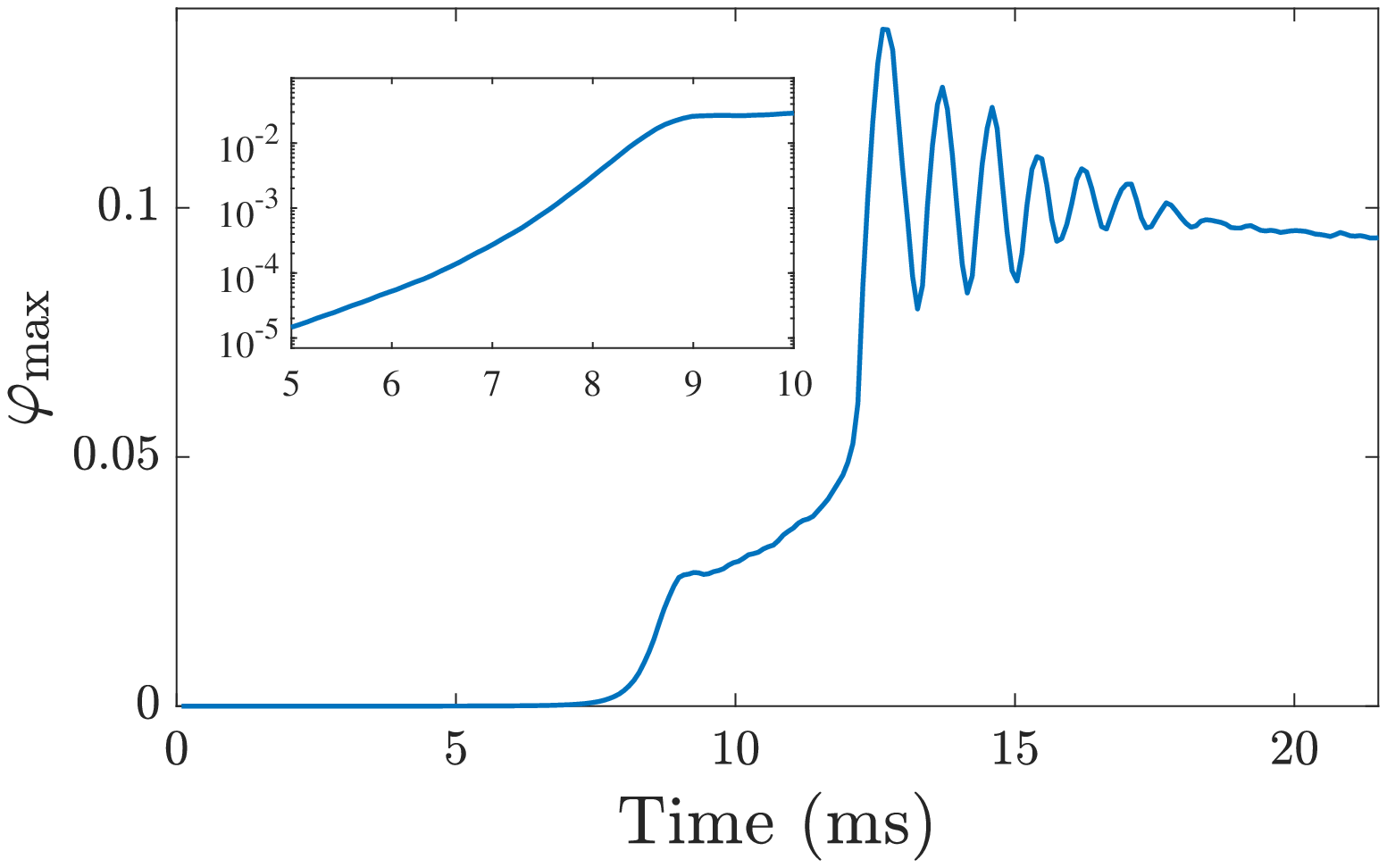}
		\caption{Maxima of the Gauss-Bonnet invariant (top) and the scalar field (bottom) for the binary Model A as functions of simulation time. The insets zoom in the evolution of both quantities in the first 11~ms. For both panels, theory parameters are set as $\beta=200$, $\lambda/m_1=8.79$, and $\iota=-1$. 
		} 
		\label{fig:4stages}
	\end{figure}
    
    A mismatch between the triangles and the filled circles in Fig.~\ref{fig:bndry} is observed for each model. This is expected especially right to each vertical line because of the fact that we do not start with correct scalar field initial data but instead with a small scalar field perturbation. Thus, until the system settles to a qusi-equlibrium configuration on a timescale dependent on $\lambda$, the initial binary separation will influence the results. What is important, though, is that the onset of the scalarization that has a significant dynamical origin (left-end of each curves) is almost independent on the initial separation with the exception of the transition region in the vicinity of the bifurcation line. 
   
    As expected, the dynamical scalarization window is larger than that for isolated NSs, and the difference between them is maximal for equal mass binary (blue markers in Fig. \ref{fig:bndry}). If the asymmetry in the mass ratio is too large the dynamical scalarization is much more difficult to develop because the  gravitational effect from the lighter NS to the heavier companion becomes less important, and thus, its threshold approaches the limit for the isolated NS (yellow markers in Fig. \ref{fig:bndry}). In addition, as the two NSs approaches, the variation in the Gauss-Bonnet invariant reads,
    \begin{align}
        \delta{\cal R}_{\rm GB}^2\propto \frac{1}{a^6} \propto f_{\rm gw}^4,
    \end{align}
    where $a$ is the separation, and the last relation comes from the Keplerian theory. The use of Keplerian orbit is a rough estimate, while capturing well the leading order effect. This variation can then be translated to a relation between the reduction in the critical coupling function and the GW frequency at the onset of dynamical scalarization, expressed as
    \begin{align}
        \delta(^{-}\lambda_{\rm bif}) \propto f_{\rm gw, onset}^2.
    \end{align}
    The tendency is plotted as dashed line in Fig.~\ref{fig:bndry} for each model.
    
To look at how the dynamical scalarization is activated by ever-increasing maximum of $\iota{\cal R}_{\rm GB}^2$ before merger, we now focus on the case with the most pronounced mutual interaction of the scalar field, viz.~model A. We plot in the top panel of Fig.~\ref{fig:4stages} the evolution of the maximum of $\iota{\cal R}_{\rm GB}^2$, which values $\lesssim 4\times10^{-3}\,\,M_\odot^{-4}$ up to the onset of merging (at $\sim12.18$~ms), while the strength of it grows by a factor of $\lesssim7$ then settles to a few hundredths shortly after merger (at $\sim18$~ms). 
	We plot in the bottom panel of Fig.~\ref{fig:4stages} the evolution of the scalar field. When $\varphi_{\rm max}$ hits $10^{-4}$ at $6.8$~ms when $\iota{\cal R}_{\rm GB}^2$ grows $3.5\%$ from the initial value.
	In addition, a scalarized NS formed aftermath the merger, where the scalar field peaks at the remnant center with a value of $\varphi_c\sim0.1$, and the scalar field possesses a multipolar structure.

	\subsection{Scalar radiation}
	
\begin{figure}
		\centering
		\includegraphics[width=\columnwidth]{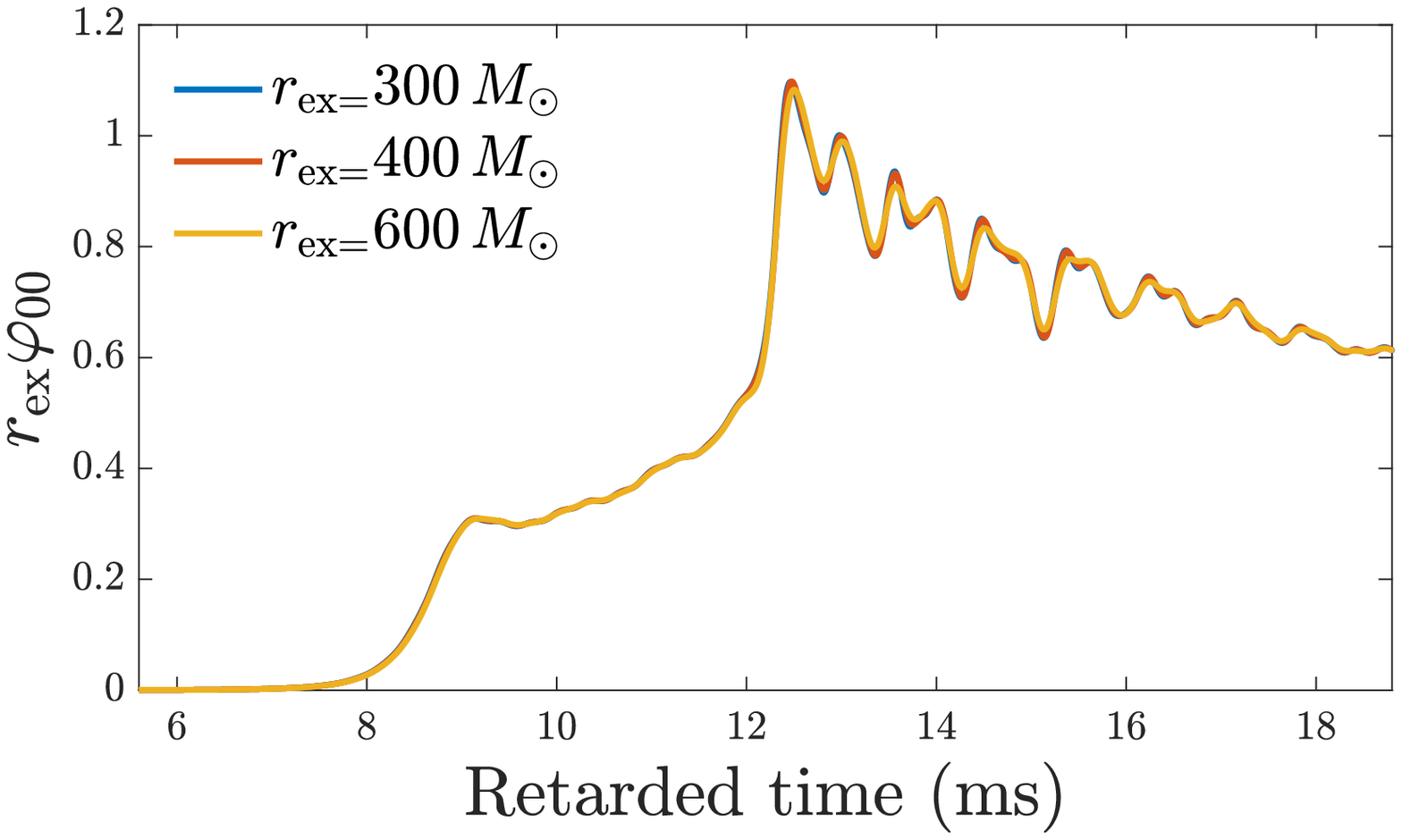}
		\includegraphics[width=\columnwidth]{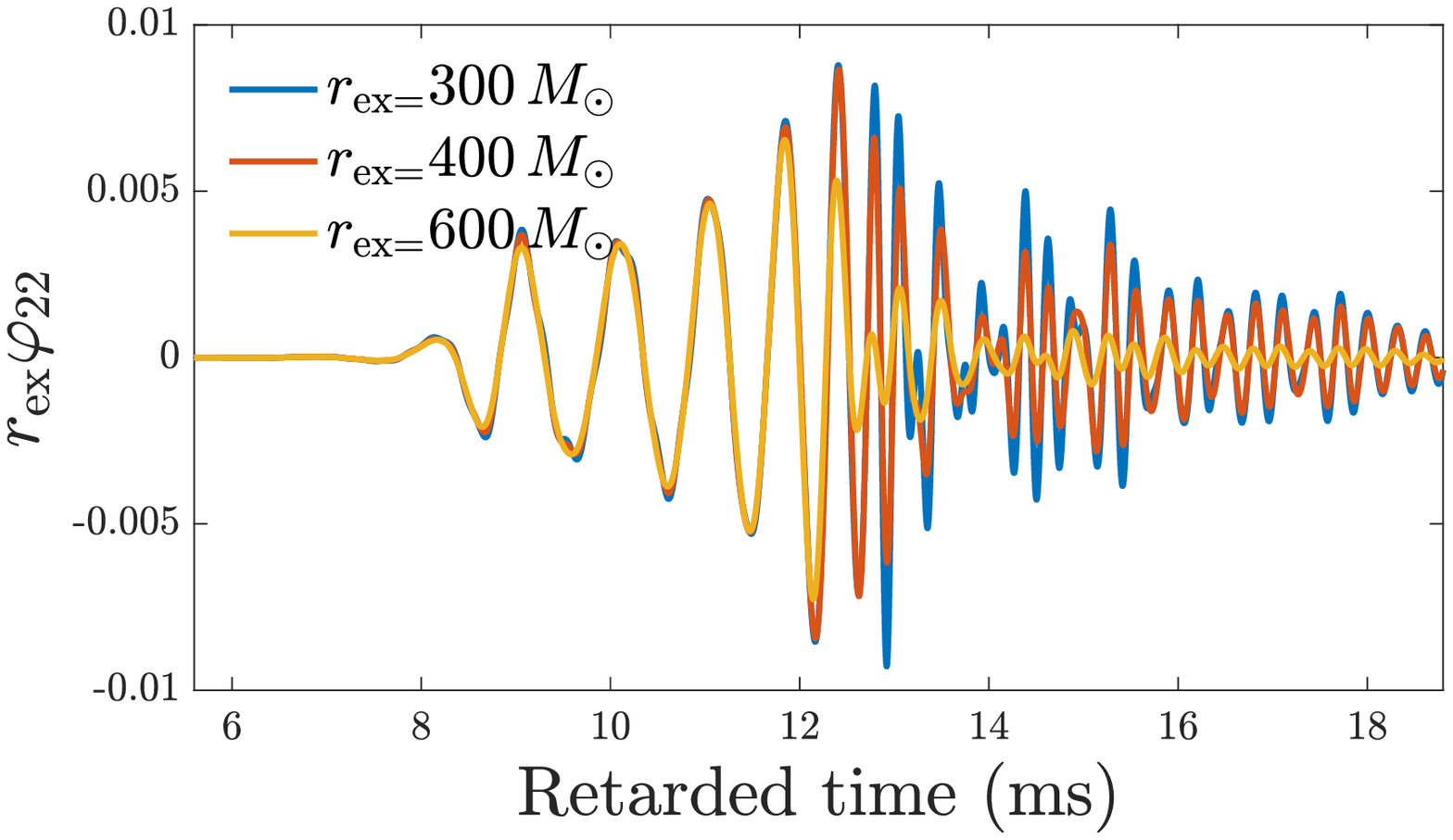}
		\caption{Monopole (top) and quadruple (bottom) moments of scalar field extracted at three different distances (in $M_\odot$): $r_{\rm ex}=300$, $400$, and $600$. 
		For both panels, the evolution is shown with respect to retarded time since the beginning of simulations, and the same binary and theory parameters as Fig.~\ref{fig:4stages} are used.
		} 
		\label{fig:phi22}
	\end{figure}

	The scalar radiation can be decomposed in the harmonic manner, namely, 
	\begin{align}\label{eq:sphi_expension}
	    \varphi(r,\theta,\phi)=\sum_{\ell m}\varphi_{\ell m}(r)\Big[{}_{0}Y_{\ell m}(\theta,\phi)\Big]
	\end{align}
	with the harmonic components defined as
	\begin{align}\label{eq:sphi_comp}
	    \varphi_{\ell m}(r)=\oint \varphi \Big[{}_{0}\overline{Y}_{\ell m}(\theta,\phi)\Big]d\Omega,
	\end{align}
    where $_{s}Y_{\ell m}$ is the spin-weighted spherical harmonics with spin $s$ and harmonic numbers $\ell$ and $m$. As relevant to the scalar-induced tensorial mode of gravitational waves, we plot in the top panel of Fig.~\ref{fig:phi22} the monopole moment of $\varphi$, denoted as $\varphi_{00}$, as a function of the retarded time with respect to the beginning of simulations. In addition, the quadruple scalar radiation, denoted as $\varphi_{22}$, is shown in the bottom panel, which kicks in after the dynamical scalarization. 
    The wavelength of the scalar-radiation during the late inspiral is a few tens of the total mass of the binary (see, e.g., Sec.~III.~C of \cite{yama08}).
    Thus, given that the total mass of the considered binaries lies in the range $2.725$--$2.892M_\odot$, an extraction radius $r_{\rm ex}=300\,\,M_\odot$ should adequately reside in the radiation zone. We extract $\varphi_{22}$ at three places in the radiation zone, where we see that the pre-merger signals have an agreement at different distances. On the other hand, the post-merger segments do not agree with different sampling distances. The reason for this is that in the present grid resolution, the grid spacing at the extraction radii is too wide to well resolve the short wavelength of post-merger scalar waves. 
    
\begin{figure}
		\centering
		\includegraphics[width=\columnwidth]{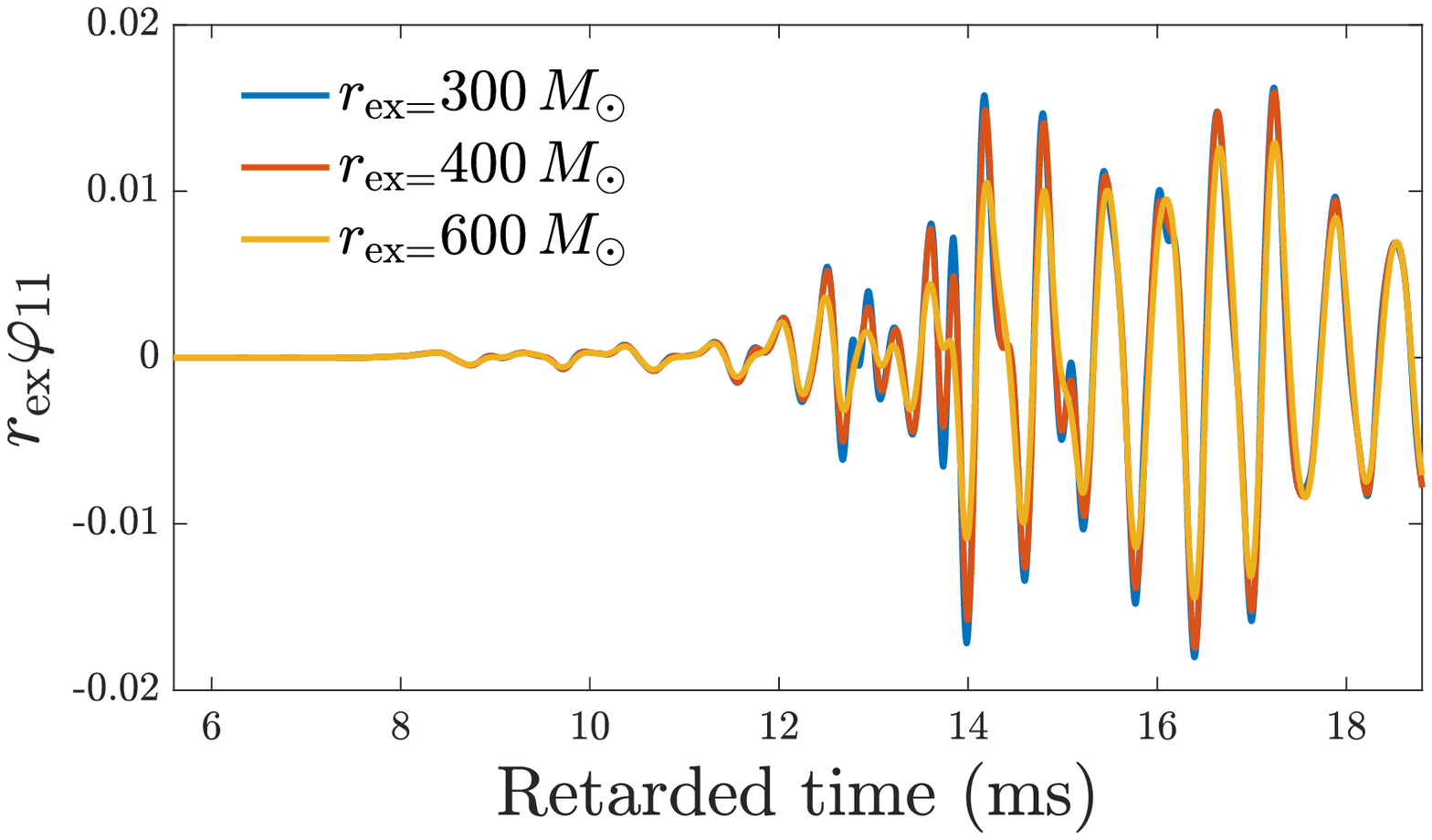}
		\includegraphics[width=\columnwidth]{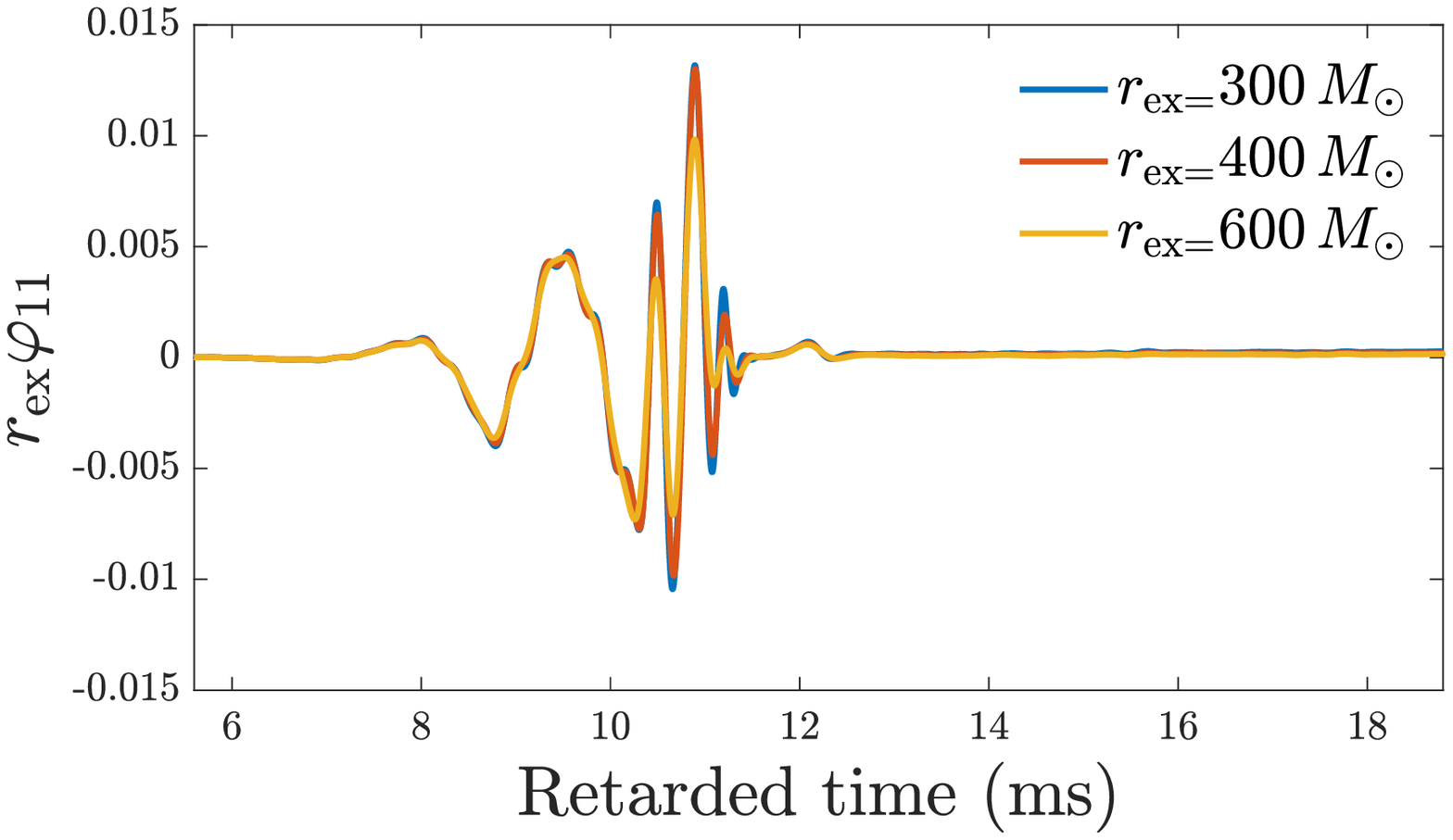}
		\caption{Dipole moment of scalar field extracted at three different distances (in $M_\odot$): $r_{\rm ex}=300$, $400$, and $600$ for model A (top) and C (bottom). 
		For both panels, the evolution is shown with respect to retarded time since the beginning of simulations, and the coupling strengths are chosen as $\lambda/m_1=8.79$ and 4.55 for the two models, respectively.
		} 
		\label{fig:dip}
	\end{figure}

    As for the leading order leakage of the orbital energy via scalar channel, we show in Fig.~\ref{fig:dip} the dipole radiation, which is supported by the difference of the scalar charge of the two NSs \cite{damo92,shir22}, for the least (model A; top panel) and the most (model C; bottom panel) asymmetric cases among the considered binaries. We see that the dipole radiation in the pre-merger phase is more significant for the latter model due to the larger difference of the scalar charges of both component. For the post-merger phase, however, we observe an abrupt quenching for the model C since a prompt collapse to BH ensues the merger. On the other hand, the hypermassive NS left behind the merger of binary A continuously emits scalar waves, and, interestingly, the dipole radiation in the post-merger stages can already be accurately traced by the used resolution, in oppose to the case of quadrupole moment, while the extraction is still not precise during merger (cf.~Fig.~\ref{fig:dip}).
	
	The more detailed study of the scalar waveform will be deferred to a future work, while we leave some comments on the implication of the absence of scalar radiation in the event GW170817. Although the constrain on the dipole radiation from GW170817 is not as stringent as that comes from pulsar binaries \cite{abbo19} with current observatories, the future detectors may have capability to push the constraints from gravitational waves to higher accuracy. The implication of the non-detection of such additional channel of energy flux than GR in a future GW170817-like systems would then set a bound of $^{-}\lambda/m_1<8$, as indicated by the result of model A (Fig.~\ref{fig:bndry}). We note that model A is the most symmetric binary among those satisfying the inferred source parameters. Thus, the most conservative constraint is obtained by considering the symmetric model.

	\section{Closing Remarks}\label{discuss}
	
	\subsection{Conclusion}
	In a subclass of SGB, spontaneous scalarization can occur for isolated NSs with a mass in a certain range, which depends on coupling strength $\lambda$ between the scalar field and the Gauss-Bonnet invariant [Eq.~\eqref{eq:quadratic}]. The considered theory comes in two flavors with each featuring a different sign of the coupling function \eqref{eq:cpling}. In particular, scalarized NSs exist for both signs $\iota=\pm1$, while hairy BHs can be realised only for the flavor $\iota=1$ \cite{done18bh,kuan21}. Focusing on NSs as members in a coalescing binary with $\iota=-1$, we numerically study the scalarization dynamically activated by the curvature of the whole system, and the convergence of the performed simulations is confirmed by running each model with three resolutions (Fig.~\ref{fig:conv}). 
	For a given EOS and stellar mass, there exists a critical coupling strength of the scalar to ${\cal R}_{\rm GB}^2$ so as to allow for a scalarized state of the NS. This threshold, however, will be reduced by an extent depending on the mass ratio of the binary [Fig.~\ref{fig:bndry}] due to the presence of a companion; in other words, a stable scalar hair can develop in the primary of binary NSs for lower values of the coupling parameters. In the presented results, the consistency of dynamical scalarization is confirmed by considering for each model two initial separations since their onsets are independent on the initial state of the binary, as expected (Fig.~\ref{fig:bndry}).

    Although gravitational-wave and pulsar timing observations have already left a narrow parameter space for a number of theories with a massless scalar field, these current constraints can be dodged by a massive scalar field with a mass as small as $10^{-16}$~eV (Sec.~\ref{massive}). As a matter of fact, even for a scalar field mass as small as $10^{-11}$~eV the effect on the structure of scalarized NSs is negligible (Fig.~\ref{fig:seq}), leaving the dynamics of the scalar field essentially the same as massless theories. In line with this fact, we here consider the theory as massless, while remarking that the results are quantitatively indistinguishable for a scalar field with up to the aforementioned mass.

    Fixing the chirp mass to that of GW170817, three mass ratios are considered, where model A is the most symmetric case (i.e, closest to the equal mass scenario), and model C is the most asymmetric possibility for the estimated mass range of each star (Tab.~\ref{tab:binaries}). The mutual interaction of the scalar field is significant for a binary with mass ratio close to 1 as suggested by the blue markers in Fig.~\ref{fig:bndry}. There, we see the minimum of $\lambda$ to allow for scalarization is sizable shifted down by the companion, while, on the other hand, the reduction in the threshold of coupling strength for model C is marginal since the weak influence of the secondary to the primary.
    
    The associated scalar radiation is computed [Eqs.~\eqref{eq:sphi_expension} and \eqref{eq:sphi_comp}] for monopole and quadrupole moments (Fig.~\ref{fig:phi22}). The scalar signal in the late inspiral phase has a frequency being twice the orbital one. After merger, the scalar field oscillates at much higher frequencies, which may attribute to the excitation of several orders of scalar modes. However, the adopted grid resolution is too crude to well capture the post-merger signal (bottom panel of Fig.~\ref{fig:phi22}). On the other hand, the resolution is already sufficient for extracting the post-merger dipole radiation (Fig.~\ref{fig:dip}), which attributes to the non-identical scalar strength after the scalarization in the two components.

    \subsection{Discussion}
    The studies above were performed for a coupling function with $\iota=-1$ that does not allow black hole scalarization in the static case, and is thus similar as a scalarization mechanism and behavior to the classical scalar-tensor theories. In addition, we have focused primarily on the change of the threshold for dynamical scalarization in comparison to the case of isolated NSs. Our preliminary results show, though, the case when  $\iota=1$ shows very interesting and distinct phenomenology. In addition, it will be interesting to explore the possibility of spin-induced scalarization of the merger remnant with $\iota=-1$. A detailed study of such phenomenology, including its implications for distinguishing between different flavors of the gravitational theory, is currently underway.

    Borrowing the experience from scalar-tensor theories, the physical condition admitting the dynamical scalarization should correlate strongly to the critical orbital frequency after which the binding energy of the binary defers from the GR value (cf.~Figs.~1 and 3 of \cite{tani15}). In other words, the quasiequilibrium sequence in the considered SGB theory will deviate from that of GR for orbital frequencies larger than the critical value. The binary will thus keep abreast of a different evolution track afterwards. Constructing quasiequilibrium sequence in SGB will be a future pursue. To that purpose, our results of the threshold for dynamical scalarization may serve to pick the most suitable parameter for such investigation.
	
	In this work, we have only considered non-rotating NS members. Although a mild stellar spin does not influence much the $M_\star$--$R_\star$ curve of scalarized NSs, a large spin can greatly modifies the profile of an equilibrium \cite{done18st}. As an example of the consequences, a rather distinct universal relation between moment of inertia and quadrupole moment is rendered \cite{done14st}. Stellar spins may therefore partially affect the onset of dynamical scalarization.
	It is also important to note that we simulated the binary evolution for massless scalar field, while a scalar mass of $m_\varphi \alt 10^{-11}$~eV will not alter the picture in any conceivable way (Sec.~\ref{massive}). Nonetheless, the waveform may be influenced by this amount of $m_\varphi$ to an extent that can deteriorate the gravitational wave analysis. A study taking into account the backreaction of scalar field in the Einstein equations with a massive scalar field is therefore necessary to investigate the waveform generated during the stages right before merger.

	\section*{Acknowledgement}
    We thank Karim Van Aelst for expert instruction on the use  of FUKA. 
	Numerical simulation was performed on Sakura cluster at Max Planck Computing and Data Facility. This work was in part supported by Grant-in-Aid for Scientific Research (Grant No.~JP20H00158) of Japanese MEXT/JSPS, and by the European Union-NextGenerationEU, through the National Recovery and Resilience Plan of the Republic of Bulgaria, project No. BG-RRP-2.004-0008-C01. DD acknowledges financial support via an Emmy Noether Research Group funded by the German Research Foundation (DFG) under grant No. DO 1771/1-1.
	
	\bibliographystyle{apsrev4-2}
	\bibliography{references}

	\appendix

\end{document}